# Explicit Commutativity Conditions for Second-order Linear Time-Varying Systems with Non-Zero Initial Conditions

Mehmet Emir KOKSAL

*Department of Mathematics, Ondokuz Mayis University, 55139 Atakum, Samsun, Turkey*

emir_koksal@hotmail.com

**Abstract:** Although the explicit commutativitiy conditions for second-order linear time-varying systems have been appeared in some literature, these are all for initially relaxed systems. This paper presents explicit necessary and sufficient commutativity conditions for commutativity of second-order linear time-varying systems with non-zero initial conditions. It has appeared interesting that the second requirement for the commutativity of non-relaxed systems plays an important role on the commutativity conditions when non-zero initial conditions exist. Another highlight is that the commutativity of switched systems is considered and spoiling of commutativity at the switching instants is illustrated for the first time. The simulation results support the theory developed in the paper.

## 1. Introduction

If Systems $A$ and $B$ are connected sequentially as shown in Figure 1, which is also known as series or cascade connection [1-4] so that $x(t) = x_A(t)$ is the input of the combined system $(A, B)$, then $y_A(t) = x_B(t)$ and $y_B(t) = y(t)$ becomes the output of $(A, B)$. Similarly is defined the cascade connection $(B, A)$. If the connections $(A, B)$ and $(B, A)$ have the same input-output relation, it is said that $A$ and $B$ are commutative.

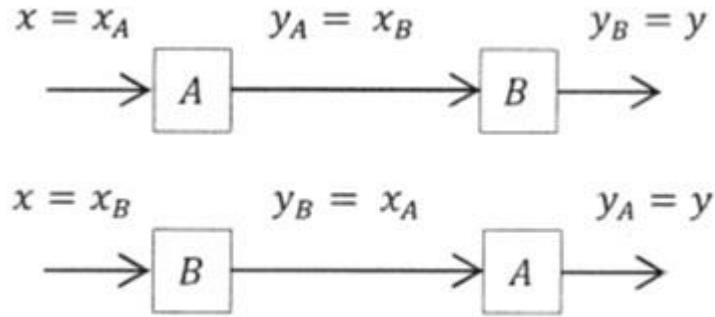

**Figure 1:** Cascade connection of the differential system *A* and *B*

Series or cascade connection of subsystems of simple orders are a very commonly used tool in the design of many engineering systems. In fact, this connection is very important for especially electrical and electronics engineers. In general, the sequence of subsystems in the series connection mainly depends on the special design approach, engineering ingenuity and traditional synthesis methods. However, performance properties such as linearity, stability, sensitivity, noise disturbance, robustness should be considered at all stages of the design so that the change of the order of connection without changing the main function of the total system (commutativity) may be more beneficial. Hence, the commutativity becomes one of the important concepts for system performance improvements in practice.

The first paper about the commutativity was studied by J. E. Marshall in 1977 [5] and he introduced the commutativity concept for the first time and investigated the commutativity of the first-order continuous-time linear time-varying systems. He proved a considerable idea that "a time-varying system can be commutative with another time-varying system only", which are very important for the further developments of the subject in the sequel. After that, commutativity conditions for second-order systems were first appeared in 1982 [6]. Then, in 1984 [7] and in 1985 [8], commutativity conditions for third and fourth-order continuous-time linear time-varying systems were studied respectively. The content of the undistributed work [8] can be found in journal paper [9] which states a comprehensive analysis on the commutativity of continuous-time linear time-varying systems. That paper presents necessary and sufficient conditions for the commutativity of relaxed systems (systems with zero initial conditions) as the first basic tutorial paper in the literature on the commutativity of continuous-time linear time-varying systems of any order. Between 1989 and 2011, no publication about commutativity has been appeared in the literature. In 2011 [10], the second basic journal publication appeared. In this paper, commutativity of non-relaxed systems (systems with non-zero initial conditions), commutativity of Euler's systems, some new results about the effects of commutativity and the explicit commutativity conditions for fifth-order systems were studied.

Commutativity property of systems has been reported to may lead many advantages from stability, sensitivity, noise robustness point of views [11-13] which are important beneficial for system design engineers [14-31].

For second-order systems which constitute a grate deal of physical and engineering applications [32-37], the explicit commutativity conditions for relaxed systems have already been derived and presented in the literature [6-9, 38-41]. However, the explicit commutativity conditions for non-relaxed second-order systems have not been studied and appeared in the literature yet; and this paper fills this vacancy. It is organized as follows:

Section 2 summarizes the results for the first set of commutativity conditions [Theorem: Koksal 1 in [10]] as already presented explicitly in the literature [9]. Section 3 includes the general form of the second set of commutativity conditions [Theorem: Koksal 2 in [10]] which are necessary and sufficient together with the first set of commutativity conditions for commutativity of non-relaxed systems. Starting from these results, the explicit commutativity conditions for second-order non-relaxed systems are derived in Section 4. Section 5 includes several consequences obtained in Section 4; these are about the cases i) the general case, ii) feedback systems, iii) one of the subsystems in $(A, B)$ is of order 1 and finally iv) one of the subsystems is a scalar system. Section 6 introduces an example to illustrate some of the obtained results. Section 7 is devoted to the commutativity conditions for switched systems, which is a subject treated for the first time in the literature. Finally, the paper ends with Section 8 which covers Conclusions.

## 2. Explicit Commutativity Conditions for Relaxed Second-order Systems

For description and modelling, analyzing, solving real engineering problems, differential equations appears in electromagnetic, electrodynamics, fluid dynamics, wave motion, wave distribution such as subfields of electric-electronics engineering and in many other sciences and branches of engineering. Especially, they are used in system and control theory that deal with the behavior of dynamical systems with inputs, and how their behavior is modified by different combinations such as cascade and feedback connections because control systems are everywhere in our life and the principles of control have a huge impact on diverse fields as engineering. When the cascade connection in system design is considered, the commutativity concept places an important role to improve different system performances. Continuous time-varying systems are modeled by ordinary differential equations though discrete time-varying systems are modeled by difference equations.

Consider two linear time-varying analog systems of second-order described by

$$A: \quad a_2(t)\ddot{y}_A(t) + a_1(t)\dot{y}_A(t) + a_0(t)y_A(t) = x_A(t), \qquad (1a)$$

$$B: \quad b_2(t)\ddot{y}_B(t) + b_1(t)\dot{y}_B(t) + b_0(t)y_B(t) = x_B(t); \qquad (1b)$$

where $x_A(t)$ and $y_A(t)$ are the input and output of system $A$, respectively; $y_A(t_0), \dot{y}_A(t_0)$ are the initial conditions at time $t=t_0 \in R$; $a_2(t) \neq 0 \;\forall t \geq t_0$; the overhead single (double) dot represents the first (second) order derivative with respect to $t$.

It is well known that [7-9], System $A$ described by Eq. (1a) be commutative with another system $B$ of the same type as expressed in Eq. (1b) under zero initial conditions, the coefficients of $B$ are expressed interms of the coefficients of $A$ by

$$\begin{bmatrix} b_2 \\ b_1 \\ b_0 \end{bmatrix} = \begin{bmatrix} a_2 & 0 & 0 \\ a_1 & a_2^{0.5} & 0 \\ a_0 & f_A & 1 \end{bmatrix} \begin{bmatrix} k_2 \\ k_1 \\ k_0 \end{bmatrix}, \; f_A = \tfrac{1}{4}[a_2^{-0.5}(2a_1 - \dot{a}_2)]; \qquad (2a)$$

where $k_2, k_1, k_0$ are constants and it must hold that

$$-a_2^{0.5} \tfrac{d}{dt}[a_0 - f_A^2 - a_2^{0.5} \dot{f}_A] k_1 = 0. \qquad (2b)$$

These are the explicit set of necessary and sufficient conditions for commutativity of relaxed second-order linear time-varying systems. They constitute the first set of necessary and sufficient conditions for non-relaxed systems reduced to explicit form for second-order relaxed systems [Theorem: Koksal1]. The problem concerned in this paper is what the commutativity conditions are if the systems $A$ and $B$ have nonzero initial conditions.

## 3. General Second Set of Commutativity Conditions for Non-Relaxed Second-order Systems

If Subsystems $A$ and $B$ are required to be commutative under non-zero initial conditions as well, a second set of conditions must be satisfied [Theorem: Koksal 2 in [9]]. These conditions are expressed for Systems $A$ of order $n$ and $B$ of order $m \leq n$, respectively, and with non-zero initial conditions.

i. First set of conditions explicitly written for 2nd order systems in Eqs. (2a) and (2d) be satisfied.
ii. Initial conditions satisfy the following at the initial time $t_0$:

$$\left\{\binom{n}{m}\begin{bmatrix} I & 0 \\ -A_2^{-1}A_1 & A_2^{-1} \end{bmatrix} - \binom{m}{n}\begin{bmatrix} 0 & I \\ B_2^{-1} & -B_2^{-1}B_1 \end{bmatrix}\right\}\begin{bmatrix} Y_A \\ Y_B \end{bmatrix} = [0]; \qquad (3a)$$

where $Y_A = [y_A(t_0) \; \dot{y}_A(t_0) \; \cdots \; y_A^{(n)}]^T$, $Y_B = [y_B(t_0) \; \dot{y}_B(t_0) \; \cdots \; y_B^{(m)}(t_0)]^T$ and the matrix $A_1$ ($A_2$, $B_1$, $B_1$) is defined by its entires $a'_{ij}$ ($a''_{ij}, b'_{ij}, b''_{ij}$) as follows:

$$a'_{ij} = \sum_{s=\max(0,i-j)}^{i-1} \frac{(i-1)!}{s!\,(i-1-s)!} a_{j-i+s}^{(s)}; \; for \; i = 1,2,\cdots,m; \; j = 1,2,\cdots,n,$$

$$a''_{ij} = \sum_{s=0}^{i-j} \frac{(i-1)!}{s!\,(i-1-s)!} a_{j-i+n+s}^{(s)}; \; for \; i = 1,2,\cdots,m; \; j = 1,2,\cdots,i,$$

$$= 0; \text{ for } i = 1,2,\cdots,m-1; \; j = i+1, i+2, \cdots, m,$$

$$b'_{ij} = \sum_{s=\max(0,i-j)}^{i-1} \frac{(i-1)!}{s!\,(i-1-s)!} b^{(s)}_{j-i+s}; \text{ for } i = 1,2,\cdots,n; \; j = 1,2,\cdots,m,$$

$$b''_{ij} = \sum_{s=\max(0,i-j-m)}^{i-j} \frac{(i-1)!}{s!\,(i-1-s)!} b^{(s)}_{j-i+m+s}; \text{ for } i = 1,2,\cdots,n; \; j = 1,2,\cdots,i,$$

$$= 0; \text{ for } i = 1,2,\cdots,n-1; \; j = i+1, i+2, \cdots, n. \tag{3b}$$

## 4. Explicit Commutativity Conditions for Non-Relaxed Second-order Systems

Although the general second set of commutativity conditions for non-relaxed systems are presented in the previous section, their application to second order systems yield to deduce more compact and useful interesting results which we call explicit commutativity conditions for second order linear time-varying systems with non-zero initial conditions. When the results in Eqs. (3a) and (3b) are simplified for $n = m = 2$, the following equations are obtained:

$$Y_B = \begin{bmatrix} Y_B(t_0) \\ \dot{Y}_B(t_0) \end{bmatrix} = \begin{bmatrix} Y_A(t_0) \\ \dot{Y}_B(t_0) \end{bmatrix} = Y_A, \tag{4a}$$

$$[A_2^{-1}(I - A_1) - B_2^{-1}(I - B_1)] \begin{bmatrix} Y_A(t_0) \\ \dot{Y}_B(t_0) \end{bmatrix} = \begin{bmatrix} 0 \\ 0 \end{bmatrix}. \tag{4b}$$

Equation 4a means that second-order systems $A$ and $B$ must have equal initial conditions, and Eq. (4b) implies that the equal initial condition vector $Y_A = Y_B$ is in the null space of $[A_2^{-1}(I - A_1) - B_2^{-1}(I - B_1)]$ at time $t_0$. Computing $A_1, A_2, B_1, B_2$ by using Eq. (3b) we have

$$A_1 = \begin{bmatrix} a_0 & a_1 \\ \dot{a}_0 & \dot{a}_1 + a_0 \end{bmatrix}, \quad A_2 = \begin{bmatrix} a_2 & 0 \\ \dot{a}_2 + a_1 & a_2 \end{bmatrix}; \tag{5a}$$

$$B_1 = \begin{bmatrix} b_0 & b_1 \\ \dot{b}_0 & \dot{b}_1 + b_0 \end{bmatrix}, \quad B_2 = \begin{bmatrix} b_2 & 0 \\ \dot{b}_2 + b_1 & b_2 \end{bmatrix}. \tag{5b}$$

Inserting Eqs. (5a, b) in (4b), using Eq. (2a), organizing and simplifying the terms, dividing the first implicit equation in (4b) by $a_2^2(t_0) \neq 0$ and the second by $a_2^{1.5}(t_0) \neq 0$, we obtain after a rigorous work:

$$\begin{bmatrix} (k_2 + k_0 - 1) + k_1 f_A & a_2^{0.5} k_1 \\ a_2^{-0.5} k_1 (1 - a_0 + a_2^{0.5} \dot{f}_A) & (k_2 + k_0 - 1) - k_1 f_A \end{bmatrix} \begin{bmatrix} Y_A(t_0) \\ \dot{Y}_A(t_0) \end{bmatrix} = \begin{bmatrix} 0 \\ 0 \end{bmatrix}. \tag{6a}$$

Hence, for commutativity with nonzero initial conditions, it is necessary and sufficient that $A$ and $B$ have equal initial conditions as in Eq. (4a) and $[Y_A(t_0) \; \dot{Y}_A(t_0)]$ be in the null space of the $2 \times 2$ coefficient matrix in Eq. 6a computed at the initial time $= t_0$.

For the existence of non-zero initial conditions which are equal for systems $A$ and $B$ due to Eq. (4a), the mentioned coefficient matrix in Eq. (6a) must be singular [42, 43], so its determinant is zero, that is

$$\Delta = (k_2 + k_0 - 1)^2 - k_1^2 + k_1^2 [a_0 - f_A^2 - a_2^{0.5} \dot{f}_A] = 0 \tag{6b}$$

at time $t_0$. It is very important and interesting as well that the term in bracket is the same term appearing in the bracket of Eq. (2b) which is the second explicit commutativity condition of the first set stated for relaxed systems.

Hence, together with the the first set of conditions in Eqs. (2a) and (2b) for relaxed systems, Eqs. (4a), and (6) set forwards additional commutativity conditions explicitly to be satisfied for second-order linear time-varying systems in case of non-zero initial conditions.

## 5. General Case and Reduction for Some Special Cases

We now consider several consequences of the above result:

**i.** Case $k_1 \neq 0$: In this case, Eq. (2b) is equivalent to write

$$a_0 - f_A^2 - a_2^{0.5} f_A = A_0, \tag{7a}$$

where $A_0$ is a constant. This equation is equivalent to say that the coefficient $a_0(t)$ of $A$ is not independant of $a_2(t)$ and $a_1(t)$ and it should be expressible in terms of them in the form

$$a_0 = A_0 + f_A^2(a_2, a_1) + a_2^{0.5} f_A(a_2, a_1). \tag{7b}$$

Then Eq. (6b) becomes

$$\Delta = (k_2 + k_0 - 1)^2 - k_1^2(1 - A_0) = 0. \tag{7c}$$

Hence, if the constants $k_2, k_1, k_0$ and $A_0$ satisfy Eq. (7c), Systems $A$ and $B$ are comutative with non-zero initial conditions

$$y_A(t_0) = y_B(t_0), \tag{8a}$$

$$y'_A(t_0) = y'_B(t_0) = -a_2^{-0.5}(t_0) \left[\frac{k_2 + k_0 - 1}{k_1} + f_A(t_0)\right] y_A(t_0),$$

$$= -\left[\frac{k_2 + k_0 - 1}{k_1} a_2^{-0.5}(t_0) + \frac{2a_1(t_0) - \dot{a}_2(t_0)}{4 a_2(t_0)}\right] y_A(t_0) \tag{8b}$$

where the second equation is obtained from Eq. (6a) by solving for $\dot{y}_A(t_0)$ and substituting value of $f_A$ from Eq. (2a) in.

**ii.** Case $k_1 = 0$: This is a case where System $B$ is a feed backed version of System $A$ with constant feedforward $\alpha_A = 1/k_2$ and feedback $\sigma_A = k_0$ gains [44-47] as shown in Fig. 2. Under zero initial conditions, it is well known that they are always commutative [8, 9]. However, when the initial conditions do exist, this case has not been considered before in [8, 9]. Eq. (4a) and Eq. (6a) with $k_1 = 0$ must be satisfied for comutativity. In fact with $k_1 = 0$, Eq. (6a) reduces to

$$\begin{bmatrix} k_2 + k_0 - 1 & 0 \\ 0 & k_2 + k_0 - 1 \end{bmatrix} \begin{bmatrix} y_A(t_0) \\ \dot{y}_A(t_0) \end{bmatrix} = \begin{bmatrix} 0 \\ 0 \end{bmatrix}. \tag{9}$$

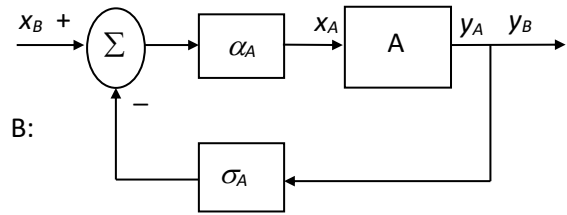

**Figure 2:** Feedback system $B$ obtained from $A$ by constant forward and feedback path gains $\alpha_A = 1/k_2, \sigma_A = k_0$, respectively.

If $k_2 + k_0 \neq 1$, this equation together with Eq. (4a) yield

$$y_A(t_0) = y_B(t_0) = 0, \tag{10a}$$

$$\dot{y}_A(t_0) = \dot{y}_B(t_0) = 0. \tag{10b}$$

So that second-order feedback conjugates do not commute with any of non-zero initial conditions unless $k_2 + k_0 = 1$. On the other hand, if $B$ is obtainable from $A$ through Eq. (2a) so that $k_2 + k_0 = 1$, then the only necessary and sufficient condition for commutativity remains Eq. (4a); that is $A$ and $B$ may have arbitrary equal initial conditions $y_A(t_0) = y_B(t_0)$, and $\dot{y}_A(t_0) = \dot{y}_B(t_0)$.

**iii.** Commutativity conditions of $A$ with a first order system $B$: Eq. (2a) in fact defines all the second or lower order commutative pairs of $A$. If $k_2 = 0$ and $k_1 \neq 0$, then $B$ is a first order system. The necessary and sufficient conditions of a second-order system ($A$) commutative with a first order system ($B$) are deducted as follows:

Eq. (2a) becomes

$$\begin{bmatrix} b_1 \\ b_0 \end{bmatrix} = \begin{bmatrix} a_2^{0.5} & 0 \\ f_A & 1 \end{bmatrix} \begin{bmatrix} k_1 \\ k_0 \end{bmatrix}; f_A = \tfrac{1}{4}\left[a_2^{-0.5}(2a_1 - \dot{a}_2)\right]. \tag{11}$$

Eq. (2b) should still stay valid. For $n = 2$ and $m = 1$, computing $A_1, A_2, B_1, B_2$ from Eq. (3b) and inserting the results in Eq. (3a), we obtain sequentially

$$A_1 = [a_0 \quad a_1], A_2 = [a_2]; B_1 = \begin{bmatrix} b_0 \\ \dot{b}_0 \end{bmatrix}, B_2 = \begin{bmatrix} b_1 & 0 \\ \dot{b}_1 + b_0 & b_1 \end{bmatrix}; \tag{12c}$$

$$\begin{bmatrix} 1 & 0 & -1 \\ -\frac{1}{b_1} & 1 & \frac{b_0}{b_1} \\ -\frac{a_0}{a_2} + \frac{b_1 + b_0}{b_1^2} & -\frac{a_1}{a_2} - \frac{1}{b_1} & \frac{1}{a_2} - \frac{b_1 + b_0}{b_1^2} b_0 + \frac{b_0}{b_1} \end{bmatrix} \begin{bmatrix} y_A(t_0) \\ \dot{y}_A(t_0) \\ y_B(t_0) \end{bmatrix} = \begin{bmatrix} 0 \\ 0 \\ 0 \end{bmatrix}. \tag{12d}$$

Note that all the matrices except $B_2$ computed for $n = 2, m = 2$ as in Eq. (5), have changed their dimensions and entries for $n = 2, m = 1$ as seen in Eq. (12). The first row of equation in Eq. (12d) clearly sets

$$y_B(t_0) = y_A(t_0). \tag{13a}$$

After substituting Eq. (13a) in the remaining two rows of Eq. (12d), pre-multiplying by $B_2$, inserting values of $b_1$ and $b_0$ computed in Eq. (11), we obtain

$$\begin{bmatrix} (k_0 - 1) + k_1 f_A & a_2^{0.5} k_1 \\ a_2^{-0.5} k_1 (1 - a_0 + a_2^{0.5} f_A) & (k_0 - 1) - k_1 f_A \end{bmatrix} \begin{bmatrix} y_A(t_0) \\ \dot{y}_A(t_0) \end{bmatrix} = \begin{bmatrix} 0 \\ 0 \end{bmatrix}. \tag{13b}$$

Obviously, it is true that this equation can be simply obtained from Eq. (6a) derived for two second-order systems by replacing $k_2 = 0$ which makes System $B$ first order.

For the commutativity with nonzero initial conditions, Eq. (13b) requires that the determinant of the coefficient matrix must be zero; that is

$$\Delta = (k_0 - 1)^2 - k_1^2 + k_1^2 [a_0 - f_A^2 - a_2^{0.5} \dot{f}_A] = 0. \tag{14a}$$

Naturally, this determinant can be obtained from Eq. (6b) by replacing $k_2 = 0$ as well. Since $k_1 \neq 0$, the term in the bracket is constant as in Eq. (7a) (due to Eq. (2b)), together with Eqs. (13a), (14a), Eq. (13b) yields that

$$\dot{y}_A(t_0) = -\left[\frac{k_0 - 1}{k_1} a_2^{-0.5}(t_0) + \frac{2a_1(t_0) - \dot{a}_2(t_0)}{4a_2(t_0)}\right] y_A(t_0). \tag{14b}$$

And this completes the conditions of commutativity with non-zero initial conditions.

As a result, in addition to Eqs. (11) and (7a) which are both valid for all $t \geq t_0$, Eq. (13a), Eqs. (14a) and (14b) which are all valid at the initial time $t_0$ constitute the conditions of commutativity of a second-order linear time-varying analog system with non-zero initial conditions with a first order system of the same type.

**iv.** Further Reduction for a Scalar system: When we choose $k_2 = 0, k_1 = 0$, the scalar (0-order) system $B$ that $A$ can be comutative is obtained from Eq. (2a) is represented by $b_0 = k_0$ constant. Hence, the only scalar system that can be commutative with a 2nd order time-varying system is a constant gain (time-invariant) system. When the initial conditions present, Eq. (3b) with $n = 2, m = 0$ yields only

$$B_2 = \begin{bmatrix} b_0 & 0 \\ \dot{b}_0 & b_0 \end{bmatrix} = \begin{bmatrix} b_0 & 0 \\ 0 & b_0 \end{bmatrix}; \tag{15a}$$

and all other matrices in Eq. (3a) are null. Hence with Eq. (15a), Eq. (3a) reduces to

$$\begin{bmatrix} 1 - b_0 & 0 \\ 0 & 1 - b_0 \end{bmatrix} \begin{bmatrix} y_A(t_0) \\ \dot{y}_A(t_0) \end{bmatrix} = \begin{bmatrix} 0 \\ 0 \end{bmatrix}. \tag{15b}$$

To satisfy this equation with arbitrary non-zero initial values $b_0 = 1$, that is system $B$ must be identity. Hence, the only scalar system that can be commutative with a second-order linear time-varying system with non-zero initial conditions is the identity (a scalar system with constant gain=1). Note that if the initial conditions are zero for the second-order system, the necessity of gain being one is redundant for commutativity.

We express the results of this section by 4 theorems in accordance to the items treated in cases i-iv.

**Theorem 1:** For the commutativity of second-order linear time-varying system $A$ described by Eq. (1a) with another second-order system $B$ which is described by Eq. (1b) ($k_2 \neq 0$) and which is not obtainable from $A$ with constant feed-forward and feed-back path gains ($k_1 \neq 0$), the necessary and sufficient conditions are that

I. Under zero initial conditions:
   i) Coefficients of $B$ must be expressible in terms of coefficients of $A$ by Eq. (2a) ($k_2 \neq 0, k_1 \neq 0$),
   ii) Coefficients of $A$ must satisfy Eq. (7a) for all t,

II. Under non-zero initial conditions:
   i) The above conditions stated for zero initial conditions are satisfied.
   ii) The constants $k_2, k_1, k_0$ in Eq. (2a) and $A_0$ in Eq. (7a) must satisfy Eq. (7c).
   iii) The initial conditions of $A$ and $B$ must satisfy Eq. (8a) and (8b).

**Theorem 2:** For the commutativity of second-order linear time-varying system $A$ described by Eq. (1a) with another second-order system $B$ which is is not obtainable from $A$ with constant feed-forward path gain $\alpha_A$ and feed-back path gain $\sigma_A$ ($k_1 \neq 0$), the necessary and sufficient conditions are that

I. Under zero initial conditions, they are always commutative unconditionally.
II. Under non-zero initial conditions:
   i) If $1/\alpha_A + \sigma_A = 1$, the system commute for all arbitrary but equal initial conditions, that is Eq. (4a) must be satisfied.
   ii) If $1/\alpha_A + \sigma_A \neq 1$, the systems do not commute with any non-zero initial conditions.

**Theorem 3:** For the commutativity of a second-order linear time-varying system $A$ described by Eq. (1a) with another linear time-varying system $B$ of first order described by Eq. (1b) with $k_2 = 0, k_1 \neq 0$, the necessary and sufficient conditions are that:

I. Under zero initial states:
   i) Coefficients of $B$ must be expressible in terms of those of $A$ as in Eq. (11).
   ii) Coefficients of $A$ must satisfy Eq. (7a) for all $t$.
II. Under non-zero initial conditions:
   i) The above conditions for zero initial states must be satisfied.
   ii) Eq. (13a) must be satisfied (equal initial conditions for outputs of $A$ and $B$).
   iii) Eq. (14b) relating the outputs of $A$ and its derivative at time $t_0$ must be satisfied.

**Theorem 4:** The only scalar system that can be commutative with a second-order time-varying system is a constant gain (time invariant) system.

## 6. Example I

Consider a system $A$ defined by

$$A: \quad 0{,}5t^2\ddot{y}_A + (t+1)\dot{y}_A + \frac{1}{2t^2}y_A = x_A \tag{16a}$$

where $a_2 = 0{,}5t^2$, $a_1 = (t+1)$, and $a_0 = 1/2t^2$ which satisfies Eqs. (2b), (7a) with $A_0 = -1/8$. When $k_2 = 2, k_1 = \sqrt{2}, k_0 = 0{,}5$, Eq. (7c) is satisfied. With these values, System $B$ is computed by Eq. (2a) as

$$B: \quad t^2\ddot{y}_B + (3t+2)\dot{y}_B + \frac{t^2+t+1}{t^2}y_B = x_B. \tag{16b}$$

$A$ and $B$ are commutative when the initial conditions are zero since all the necessary conditions are satisfied. In fact, for an input $40\sin(2\pi ft)$ with $f = 2Hz$, both systems $AB$ and $BA$ yield the same response as seen in Fig. 3 (see curve $AB = BA$: relaxed).

Further, Eqs. (8a) and (8b) require

$$y_B(t_0) = y_A(t_0), \tag{17a}$$

$$\dot{y}_B(t_0) = \dot{y}_A(t_0) = -\left[\frac{t_0^2 + 2t_0 + 3}{2t_0}\right] y_A(t_0). \tag{17b}$$

for commutativity of $A$ and $B$ under non-zero initial conditions as well. For $t_0 = 1, y_B(1) = y_A(1) = 1$, Eq. (17b) yields $\dot{y}_B(1) = \dot{y}_A(1) = -3$ with the same sine wave input, $AB$ and $BA$ yield equal response also seen in Fig. 3 (see curve $AB = BA$: unrelaxed).

Since the additional conditions for commutativity required for unrelaxed systems depend on the initial time as implied by Eq. (14b), more specifically Eq. (17b) fort his example, if the initial time $t_0$ is changed to $t_0 = 1{,}5$, the commutativity with non-zero initial conditions will be spoiled whilst in relaxed case $A$ and $B$ are still commutative; this is shown in Fig. 4 (see $AB$: not relaxed, $BA$: not relaxed, $AB = BA$ relaxed).

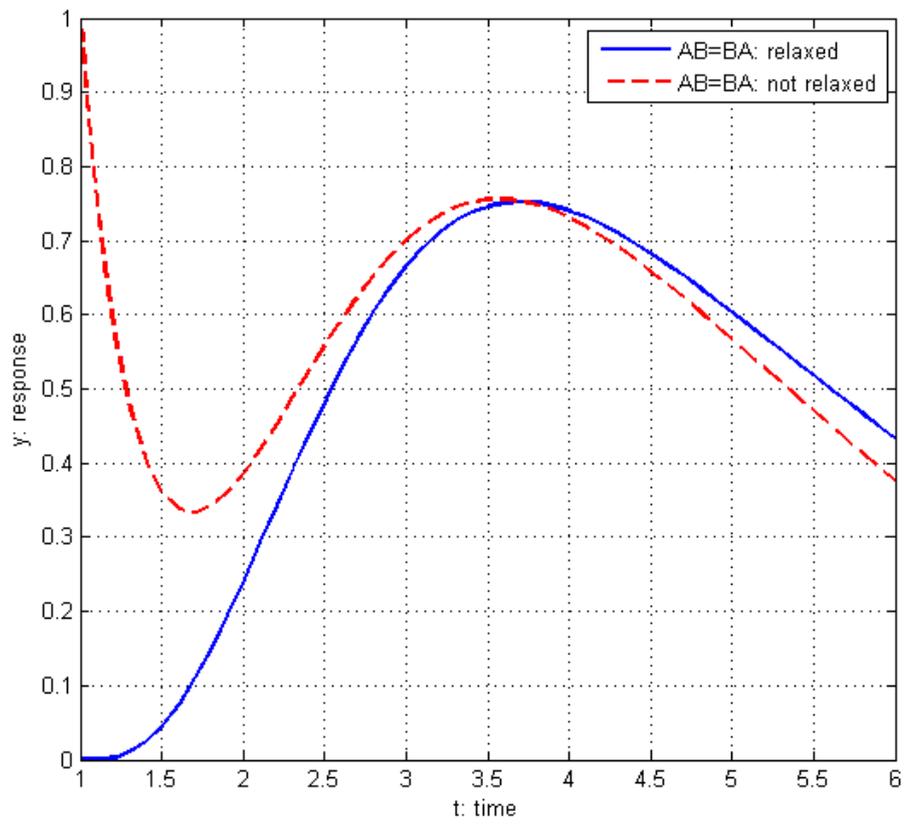

**Figure 3:** Response for *AB* with initial time $t_0 = 1$

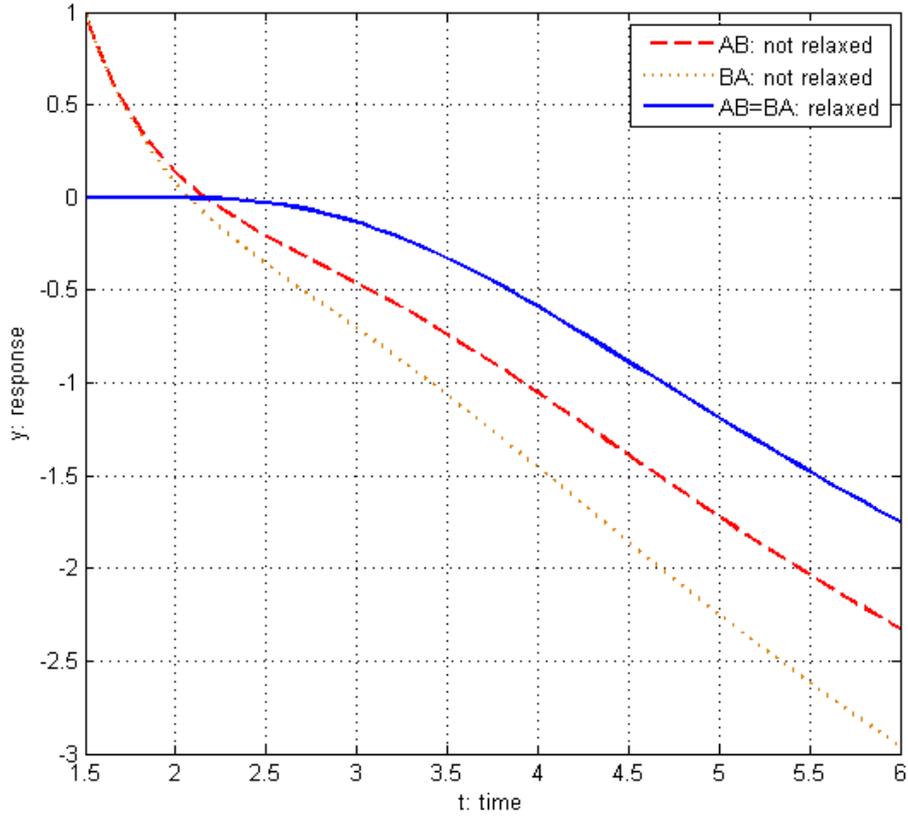

**Figure 4:** Response for *BA* with initial time $t_0 = 1,5$

## 7. Commutativity of Switched Systems

Switched systems provides mathematical modelling of many physical or man-made systems displaying switching features as in power electronics, flight and network control, switched capacitor circuits. Due to their numerous applications, many results on analysis, properties, control of switched systems have been studied in [48, 49]. Some of this literature is for nonlinear switched systems [50]. As for any system, stability is a crucial problem for switched systems as well, see the reference [51]; since this manuscript is restricted to linear time-varying systems, we consider the type of switched systems considered there in.

For a complete definition of behavior of a given linear dynamical time-varying system described by Eqs. (1a,b), it is sufficient that the coefficients of the differential equations are piecewise continuous functions. In other words, the output solutions are unique and continuous for any piecewise continuous input function [52]. Hence in the scope of switched systems considered in [51], the coefficients of these equations are piece-wise constant and changes their values at each switching time. Thus, switching parameters do not cause any discontinuities in the solutions for the intermediate output and the final output when the systems are connected in cascade. Therefore, the switching phenomenon des not contribute any discontinuity that should be taken into special account. Further, the in-differentiability of the coefficients at the switching instants do not cause any problem from the unique and continuous solvability point of views.

Although the switching of the parameters do not affect the existence and uniqueness properties of the solutions, as far as the commutativity conditions are concerned we attract attention to the followings points which have not been considered in the literature before: All the time $t \geq t_0$ except the switching instants $t_j > t_0, j = 1,2,...$ that may occur at most finite number of times in every finite interval occurring in $[t_0, \infty)$, the usual commutativity conditions are valid with the remark that the same constants $k_2, k_1, k_0$ should be valid for all switching periods. On the other hand, some of these conditions such as Eqs. (1a,b), (6a, b), (8b) involve the derivatives of the coefficients which may change abruptly due to switching and hence their derivatives is undefined or get unbounded. In this case, the commutativity conditions get violated at the switching instants and commutativity may get spoiled and become invalid after the first switching instant on. The following examples over light some of these properties for switched systems by paying attention on to the switching instants.

**Example II**

Consider the second order subsystem $A$ defined in Eq. (1a) with

$$a_2(t) = 1, \ a_1(t) = -1 + \sigma(t), \ a_0(t) = -2 + 2\sigma(t), \tag{18a}$$

$$where \ \sigma(t) = \begin{cases} 0 & for \ t \in T_u = [0,1) \cup [3,4.5) \\ 10 & for \ t \in T_s = [1,3) \cup [4.5, \infty) \end{cases}. \tag{18b}$$

Above, $T_u$ and $T_s$ represent unstable (eigenvalues are $-1$ and $+2$) and stable (eigenvalues are $-3$ and $-6$) dwelling periods. This switching signal has mode-dependent average dwell time (MDADT) property so that $A$ is designed to be exponentially stabilized [51]. Choosing

$$k_2 = 1, \ k_1 = -2, \ k_0 = 4. \tag{19}$$

and using Eq. (1a) the candidate commutative system $B$ is obtained as

$$b_2(t) = 1, \ b_1(t) = -3 + \sigma(t), \ b_0(t) = 3 + \sigma(t). \tag{20}$$

Choosing the initial condition at $t_0 = 0$ as $y_A(0) = 0.6$, and the other as to satisfy the commutativity conditions in Eqs. (8a,b) we have

$$y_B(0) = y_A(0) = 0.6, \tag{21a}$$

$$y'_A(t_0) = y'_B(t_0) = 1.5 \,. \tag{21b}$$

The cascaded systems are simulated for zero state (by applying a sinusoidal input with amplitude of -10 and frequency of 0.5 Hz with zero initial conditions) and zero input responses (by assuming the above initial conditions). The results of the outputs are shown in Fig. 5. The complete response is the summation of initial condition response and the zero state response due to linearity; therefore it is not shown in the plots. It is seen that $AB$ and $BA$ have the same forced responses (AB-forced r., BA-forced r.) until the first switching time $t = 1$, and then after they have different outputs. This is expected since until $t = 1$ all the commutativity conditions for relaxed systems are satisfied; however at $t = 1$ Eq. 1b is not satisfied due to the unboundedness of the derivative of $a_1(t)$. From this observation we drive the following conclusion: Even the commutativity conditions are satisfied for all continuous time intervals between switching instants, their violation at the switching instants due to unbounded derivatives of stepwise jumping parameters with switching may spoil commutativity.

It is true that the initial condition responses of $AB$ (AB-initial c. r.) and $BA$ (BA-initial c. r.) are not the same, This is not mainly due to switching but it is because the commutativity condition in Eq. (6b) is not satisfied.

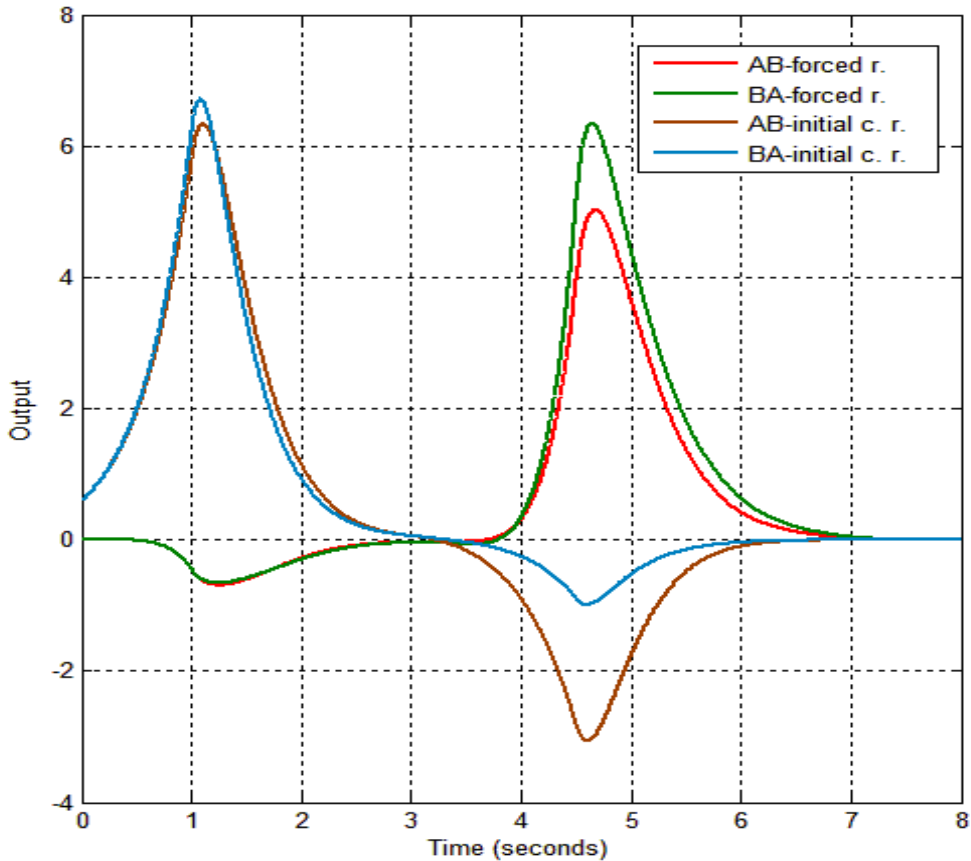

**Figure 5**: Outputs of cascade connections $AB$ and $BA$ for $k_2 = 1, k_1 = -2 \ k_0 = 4$.

To satisfy the first set of commutativity conditions, especially Eq. (1b), it is chosen that $k_1 = 0$ whilst the others are kept the same, that is $k_2 = 1, k_0 = 4$. The new subsystem $B$ is found by using Eq. (1a) as

$$b_2(t) = 1, \ b_1(t) = -1 + \sigma(t), \ b_0(t) = 2 + 2\sigma(t). \tag{22}$$

It is now satisfied all the sufficient conditions for the commutativity of $A$ and $B$ with zero initial states. See Figure 6 so that $AB$ and $BA$ have the same forced responses (AB-forced r. = BA-forced r.) which are obtained by the same input as before. However, with the same nonzero initial conditions the initial condition responses (AB-initial c. r., BA-initial c. r.) show that commutativity is invalid, the outputs are not equal at all. And this is because Eq. (6b) is not satisfied and therefore commutativity does not hold with any nonzero initial conditions as implied by Eqs (13).

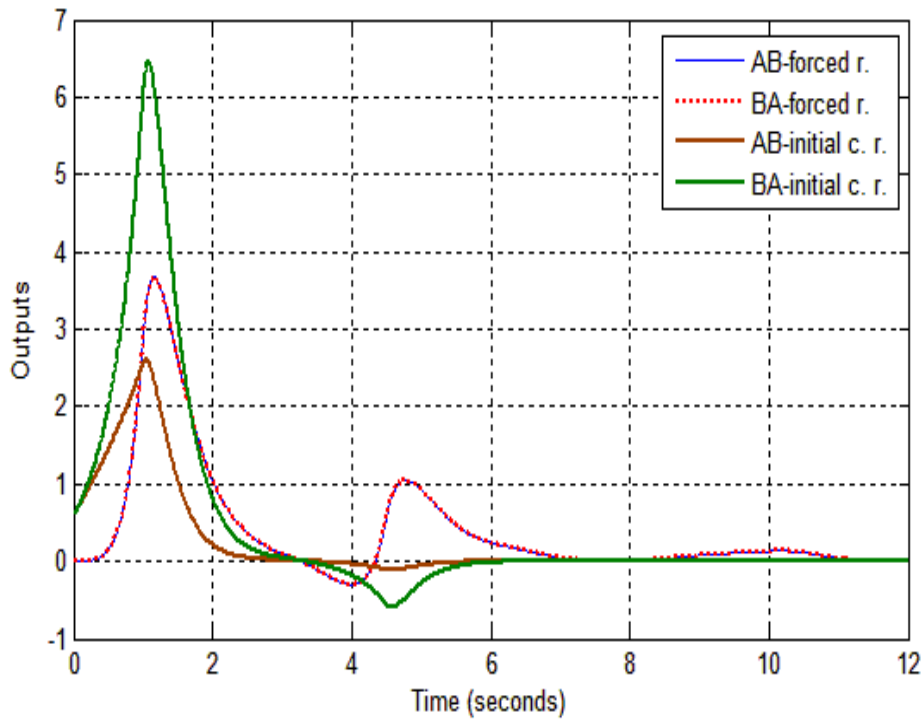

**Figure 6:** Outputs of cascade connections $AB$ and $AB$ for $k_2 = 1, k_1 = 0 \ k_0 = 4$.

**Example III**

Since the commutativity of switched systems has never been studied before, this example is considered by a different set of parameters to light the role of the commutativity condition given in (Eq. 1b). Consider the second order subsystem $A$ defined in Eq. (1a) with

$$a_2(t) = 1, \ a_1(t) = -1 + 3\sigma(t-1), \ a_0(t) = -1 + 6\sigma(t-1), \tag{18a}$$

$$where \ \sigma(t) = \begin{cases} 0 & for \ t < 0 \\ 3 & for \ t \geq 0 \end{cases}. \tag{18b}$$

Choosing

$$k_2 = 1, \ k_1 = 2, \ k_0 = 3, \tag{19}$$

and using Eq. (1a) the commutative system $B$ is obtained as

$$b_2(t) = 1, \ b_1(t) = 1 + 3\sigma(t), \ b_0(t) = 1 + 8\sigma(t). \tag{20}$$

Commutativity conditions in Eq. (2) are satisfied until $t = 1$. Hence the forced responses of $AB$ and $AB$, which are obtained by a sinusoidal input of amplitude 15 and frequency of 0.5 Hz, are identical until this time (see AB-forced r., BA-forced r. in Fig. 7). Choosing the initial condition at $t_0 = 0$ as $y_A(0) = -1$, and the others as to satisfy the commutativity conditions in Eqs. (8a,b) we have

$$y_B(0) = y_A(0) = 1, \tag{21a}$$

$$y'_A(t_0) = y'_B(t_0) = -1. \tag{21b}$$

The results for initial condition responses plotted in Fig. 7 (AB-initial c. r., BA-initial c. r.) show that they are equal only until $t = 1$. This is expected since the commutativity condition in Eq. (1b) is satisfied until this time and it is not satisfied at $t = 1$ due to switching. So commutativity is not valid in the rest of the time at all.

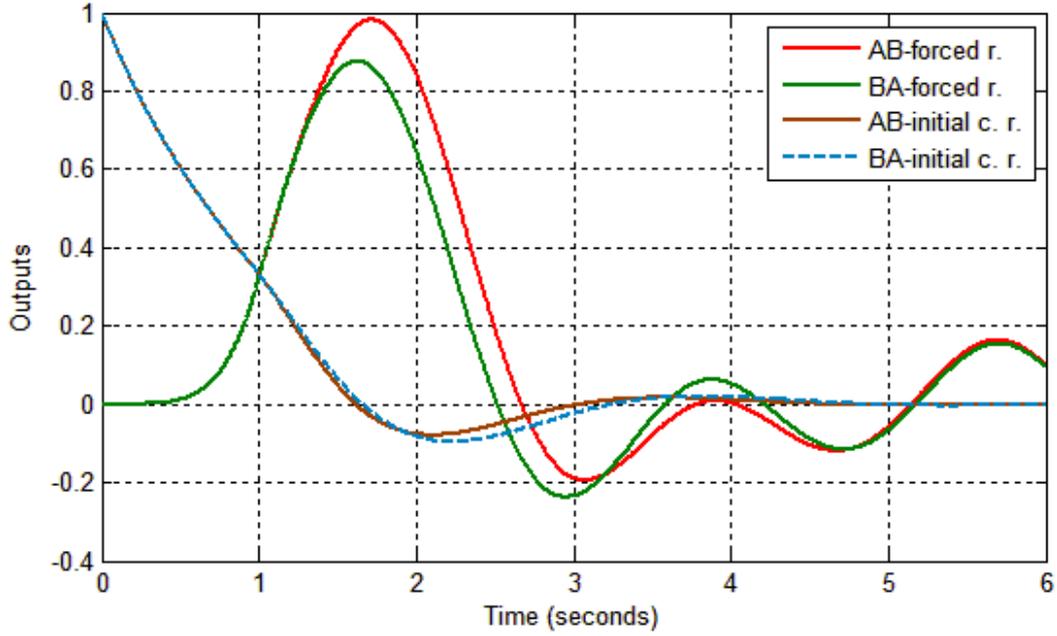

**Figure 7:** Outputs of cascade connections $AB$ and $AB$ for $k_2 = 1, k_1 = 2\ k_0 = 3$.

## 8. Conclusions

Complete set of necessary and sufficient conditions for commutativity of linear second-order time-varying systems with non-zero initial conditions are presented in the most explicit form. It is shown that the second requirement for the commutativity of relaxed systems plays an important role on the commutativity conditions when initial conditions are not zero as well. Important results have been proven for case where one of the systems is feedback version of the other, or a first order system or a scalar system. Since most of the systems encountered in the literature and engineering applications [32-37] are of second-order, the contribution of this paper seems worthy. The simulation results well verify the results obtained in the paper.

**Acknowledgments:** This paper is supported by the Scientific and Technological Research Council of Turkey under the project no. 115E952.